\documentclass[aps,reprint,prl,superscriptaddress,amsmath,amssymb,longbibliography]{revtex4-1}
\usepackage{graphicx}
\usepackage{amsmath}
\usepackage{bm}

\usepackage{color}

\definecolor{CLBlue}{rgb}{0, .3, .6}

\begin{document}

\title{Exactly solvable statistical physics models for large neuronal populations}

\author{Christopher W.~Lynn}
\email{Corresponding author: christopher.lynn@yale.edu}
\affiliation{Initiative for the Theoretical Sciences, The Graduate Center, City University of New York, New York, NY 10016, USA}
\affiliation{Joseph Henry Laboratories of Physics, Princeton University, Princeton, NJ 08544, USA}
\affiliation{Department of Physics, Quantitative Biology Institute, and Wu Tsai Institute, Yale University, New Haven, CT 06520, USA}
\author{Qiwei Yu}
\affiliation{Lewis--Sigler Institute for Integrative Genomics, Princeton University, Princeton, NJ 08544, USA}
\author{Rich Pang}
\affiliation{Princeton Neuroscience Institute, Princeton University, Princeton, NJ 08544, USA}
\author{William Bialek}
\affiliation{Joseph Henry Laboratories of Physics, Princeton University, Princeton, NJ 08544, USA}
\affiliation{Lewis--Sigler Institute for Integrative Genomics, Princeton University, Princeton, NJ 08544, USA}
\affiliation{Center for Studies in Physics and Biology, Rockefeller University, New York, NY 10065 USA}
\author{Stephanie E.~Palmer}
\affiliation{Department of Organismal Biology and Anatomy, University of Chicago, Chicago, IL 60637, USA}
\affiliation{Department of Physics, University of Chicago, Chicago, IL 60637, USA}

\date{\today}

\begin{abstract}
Maximum entropy methods provide a principled path connecting measurements of neural activity directly to statistical physics models, and this approach has been successful for populations of $N\sim 100$ neurons. As $N$ increases in new experiments, we enter an undersampled regime where we have to choose which observables should be constrained in the maximum entropy construction. The best choice is the one that provides the greatest reduction in entropy, defining a ``minimax entropy'' principle. This principle becomes tractable if we restrict attention to correlations among pairs of neurons that link together into a tree; we can find the best tree efficiently, and the underlying statistical physics models are exactly solved. We use this approach to analyze experiments on $N\sim 1500$ neurons in the mouse hippocampus, and show that the resulting model captures the distribution of synchronous activity in the network.
\end{abstract}

\maketitle

It has long been hoped that neural networks in the brain could be described using concepts from statistical physics \cite{wiener_58,cooper_73,little_74,hopfield1982,amit_89,hertz+al_91}. More recently, our ability to explore the brain has been revolutionized by techniques that record the electrical activity from thousands of individual neurons, simultaneously 
\cite{segev+al2004,litke+al2004,chung2019high,dombeck2010functional,tian2012neural,demas+al2021,steinmetz2021neuropixels}. Maximum entropy methods connect these data to theory, starting with measured properties of the network and arriving at models that are mathematically equivalent to statistical physics problems \cite{Schneidman-01,Nguyen-01}. In some cases these models provide successful, parameter--free predictions for many detailed features of the neural activity pattern \cite{Meshulam-17, Meshulam-02}. The same ideas have been used in contexts ranging from the evolution of protein families to ordering in flocks of birds \cite{Lezon-01,weigt+al_09,marks2011protein,lapedes2012using,Bialek-01,russ2020evolution,Lynn-04}. But as experiments probe systems with more and more degrees of freedom, the number of samples that we can collect typically does not increase in proportion. Here we present a strategy for building maximum entropy models in this undersampled regime, and apply this strategy to data from 1000+ neurons in the mouse hippocampus. 

Consider a system of $N$ variables $\bm{x} = \{ x_{ i}\}$, $i = 1,\, 2,\, \cdots ,\, N$. To describe the system, we would like to write down the probability distribution $P(\bm{x})$ over these microscopic degrees of freedom, in the same way that we write the Boltzmann distribution for a system in equilibrium. In the example that we discuss below, all the $x_{ i}$ are observed at the same moment in time, but we could also include time in the index $i$, so that $P(\bm{x})$ becomes a probability distribution of trajectories.

From these $N$ variables we can construct a set of $K$ operators or observables $\{ f_\nu(\bm{x})\}$, $\nu = 1,\, 2,\, \cdots ,\, K$. The maximum entropy approach takes this limited number of observables seriously, and insists that expectation values for these quantities predicted by the model match the values measured in experiment,
\begin{equation}
\left< f_\nu (\bm{x})\right>
_P = \left< f_\nu (\bm{x})\right>_\text{exp},
\label{match1}
\end{equation}
or more explicitly,
\begin{equation}
\sum_{\bm{x}} P(\bm{x}) f_\nu(\bm{x}) = {1\over M}\sum_{m=1}^M  f_\nu(\bm{x}^{(m)}) ,
\end{equation}
where $\bm{x}^{(m)}$ is the $m^\text{th}$ sample out of $M$ samples in total. Notice that to have control over errors in the measurement of all $K$ expectation values, we need to have $K \ll MN$.

There are infinitely many distributions that obey these matching conditions, but the idea of maximum entropy is that we should choose the one that has the least possible structure, or equivalently generates samples that are as random as possible while obeying the constraints in Eq.~(\ref{match1}). From Shannon we know that ``as random as possible'' translates uniquely to finding the distribution that has the maximum entropy consistent with the constraints \cite{shannon1948mathematical,Jaynes-01}. The solution of this optimization problem has the form
\begin{equation}
P(\bm{x}) = {1\over Z}\exp\left[- \sum_{\nu =1}^K \lambda_\nu f_\nu(\bm{x})\right],
\label{model1}
\end{equation}
where the coupling constants $\lambda_\nu$ must be chosen to satisfy Eq.~(\ref{match1}). We emphasize that in using this approach, the model in  Eq.~(\ref{model1}) is something that needs to be tested---in systems such as networks of neurons, there is no H--theorem telling us that entropy will be maximized, nor is there a unique choice for the constraints that would correspond to the Hamiltonian of an equilibrium system.

Without a Hamiltonian, how should we choose the observables $f_\nu$? Probability distributions define a code for the data \cite{shannon1948mathematical} in which each state $\bm{x}$ is mapped to a code word of length 
\begin{equation}
\ell(\bm{x})  = - \log P(\bm{x}) ,
\end{equation}
so the mean code length for the data is
\begin{equation}
\langle \ell \rangle_{\rm exp} \equiv  -{1\over M}\sum_{n=1}^M \log P(\bm{x}^{(n)}).
\end{equation}
Combining Eqs.~(\ref{match1}) and (\ref{model1}), $\langle \ell \rangle_{\rm exp}$ is exactly the entropy of $P$. Thus, among all maximum entropy distributions, the one that gives the shortest description of the data is the one with minimum entropy. This ``minimax entropy'' principle was discussed 25 years ago \cite{Zhu-01}, but has attracted relatively little attention. 

Every time we add a constraint, the maximum possible entropy is reduced \cite{Schneidman-02}, and this entropy reduction is an information gain. Thus the minimax entropy principle tells us to choose observables whose expectation values provide as much information as possible about the microscopic variables. The problem is that finding these maximally informative observables is generally intractable.

To make progress we restrict the class of observables that we consider. With populations of $N\sim 100$ neurons, it can be very effective to constrain just the mean activity of each neuron and the correlations among pairs \cite{Schneidman-01,Meshulam-17,Meshulam-02}. But this corresponds to $K\propto N^2$ constraints, and at large $N$ we will violate the good sampling condition $K \ll NM$. To restore good sampling, we try constraining only $N_c$ of the correlations, which define links between specific pairs of neurons that together form a graph $\mathcal{G}$. If we describe the individual neurons as being either active or silent, so that $x_i \in \{0, 1\}$, then the maximum entropy distribution is an Ising model on the graph $\mathcal{G}$,
\begin{equation}
\label{eq_PG}
P_\mathcal{G}(\bm{x}) = \frac{1}{Z}\exp\Bigg[\sum_{{(ij)}\in \mathcal{G}}J_{ij}x_ix_{ j} + \sum_i h_i x_i\Bigg] ,
\end{equation}
where $\{h_i, J_{ij}\}$ must be adjusted to match the measured expectation values $\langle x_i\rangle_{\rm exp}$ and $\langle x_ix_{ j}\rangle_{\rm exp}$ for pairs ${(ij)} \in \mathcal{G}$. The minimax entropy principle tells us that we should find the graph $\mathcal{G}$ with a fixed number of links $N_c$ such that $P_\mathcal{G}(\bm{x})$ has the smallest entropy while obeying these constraints. This remains intractable.

\begin{figure}[t]
\centering
\includegraphics[width = \columnwidth]{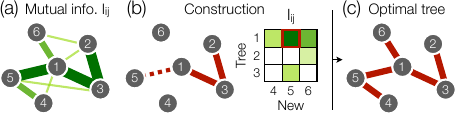} \\
\caption{Computing the optimal tree of pairwise correlations. (a) Mutual information between variables in a hypothetiacal system. (b) One can build the optimal tree one variable at a time, from an arbitrary starting variable, by iteratively adding the connection corresponding to the largest mutual information (dashed) from the tree (solid) to the remaining variables; this is Prim's algorithm. (c) Optimal tree that minimizes the entropy $S_\mathcal{T}$ and maximizes the information $I_\mathcal{T}$. \label{fig_tree}}
\end{figure}

Statistical physics problems are hard because of feedback loops. If we can eliminate these loops then we can find the partition function exactly, as in one--dimensional systems or on Bethe lattices \cite{sethna_06}. If the graph $\mathcal{G}$ has no loops then it describes a tree $\mathcal{T}$, and among other simplifications we can write the entropy 
\begin{equation}
\label{eq_IT}
-\sum_{\bm{x}} P_{\mathcal{T}}(\bm{x}) \log P_{\mathcal{T}}(\bm{x}) \equiv S_\mathcal{T} = S_{\rm ind} - \sum_{(ij)\in \mathcal{T}} I_{ij},
\end{equation}
where $S_{\rm ind}$ is the independent entropy of the individual variables, and $I_{ij}$ is the mutual information between $x_i$ and $x_j$ [Fig.~\ref{fig_tree}(a)]. Because of the constraints on the maximum entropy distribution, $I_{ij}$ is the same whether we compute it from the model or from the data. Thus we can compute the entropy of the pairwise maximum entropy model on any tree without constructing the model itself.

Minimizing the entropy $S_\mathcal{T}$ in Eq.~(\ref{eq_IT}) is equivalent to maximizing the total mutual information 
\begin{equation}
I_\mathcal{T} = \sum_{(ij)\in\mathcal{T}} I_{ij}.
\end{equation}
This defines a minimum spanning tree problem \cite{Nguyen-01}, which admits a number of efficient solutions. For example, one can grow the optimal tree by greedily attaching the new variable $x_i$ with the largest mutual information $I_{ij}$ to an existing variable $x_j$ [Figs.~\ref{fig_tree}(b) and \ref{fig_tree}(c)]; this is Prim's algorithm, which runs in $O(N^2)$ time \cite{Moore-01}. By restricting observables to pairwise correlations that form a tree, we can solve the minimax entropy problem exactly, even at very large $N$. Further, we can give explicit expressions for the fields and couplings \cite{Chow-01, Nguyen-01},
\begin{align}
\label{eq_J}
J_{ij} &= \ln \left[\frac{\langle x_i x_j\rangle \left(1 - \langle x_i \rangle -  \langle x_j \rangle +\langle x_i x_j\rangle \right)}{\left(  \langle x_i \rangle - \langle x_i x_j\rangle \right)\left(\langle x_j \rangle - \langle x_i x_j\rangle\right)}\right], \\
\label{eq_h}
h_i &= \ln \frac{ \langle x_i \rangle}{1 -  \langle x_i \rangle} \\
& \quad\quad + \sum_{j \in {\cal N}_i} \ln \left[\frac{\left(1 -  \langle x_i \rangle\right)\left( \langle x_i \rangle - \langle x_i x_j\rangle\right)}{ \langle x_i \rangle\left(1 -  \langle x_i \rangle - \langle x_j \rangle + \langle x_i x_j\rangle\right)}\right], \nonumber
\end{align}
where ${\cal N}_i$ are the neighbors of $i$ on the tree. The total number of constraints is $K = 2N-1$, so if the number of independent samples $M \gg 2$, then we are well sampled.

\begin{figure}[t]
\centering
\includegraphics[width = \columnwidth]{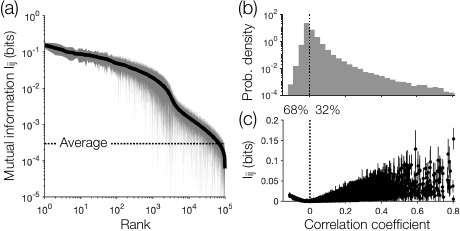} \\
\caption{Mutual information in a large population of neurons. (a) Ranked order of all significant mutual information $I_{ij}$ in a population of $N=1485$ neurons in the mouse hippocampus \cite{Gauthier-01}. Solid line and shaded region indicate estimates and errors (two standard deviations) of $I_{ij}$. (b) Distribution of correlation coefficients over neuron pairs, with percentages indicating the proportion of positively and negatively correlated pairs. (c) Mutual information $I_{ij}$ versus correlation coefficient, where each point represents a distinct neuron pair. Estimates and errors are the same in (a). \label{fig_data}}
\end{figure}

As emphasized above, our approach yields a model that needs to be tested. At large $N$, pairwise correlations are a vanishingly small fraction of all possible correlations, and in building a tree we keep only a vanishingly small fraction of these. Does this literal backbone of correlation structure contain enough information to capture something about the behavior of the network as a whole?

We analyze data from an experiment on the mouse hippocampus \cite{Gauthier-01}. Mice are genetically engineered so that neurons express a protein whose fluorescence is modulated by calcium concentration, which in turn follows the electrical activity of the cell. Recording electrical activity is then a problem of imaging, which is done with a scanning two--photon microscope as the mouse runs in a virtual environment. The fluorescence signal from each cell consists of a relatively quiet background interrupted by short periods of activity, providing a natural way to discretize into active/silent ($x_i = 1/0$) in each video frame \cite{Meshulam-17}. Images are collected at $30\,\text{Hz}$ for $39\,\text{min}$, and the field of view includes $N = 1485$ neurons. This yields $M\sim 7\times 10^4$ (non--independent) samples, sufficient to estimate the mutual information $I_{ij}$ with small errors.

Among all $N(N-1)/2 \sim 10^6$ pairs of neurons, only $9\%$ exhibit significant mutual information $I_{ij}$ [Fig.~\ref{fig_data}(a)]. We see that the distribution of mutual information is heavy--tailed, such that a small number of correlations contain orders of magnitude more information than average ($\bar{I} = 2.9\times 10^{-4}\,\text{bits}$). Additionally, while most pairs of neurons are negatively correlated [Fig.~\ref{fig_data}(b)], the strongest mutual information belong to pairs that are positively correlated [Fig.~\ref{fig_data}(c)]. Together, these observations suggest that a sparse network of positively correlated neurons may provide a large amount of information about the collective neural activity.

\begin{figure}[b]
\centering
\includegraphics[width = \columnwidth]{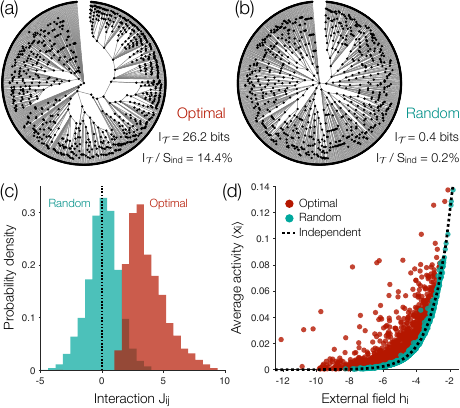} \\
\caption{Minimax entropy models of a large neuronal population. (a) The optimal tree for the neurons in Fig.~\ref{fig_data}, and (b) a random tree over the same neurons. In both trees, the central neuron has the largest number of connections, and those on the perimeter are leaves (having one connection) with distance from the central neuron decreasing in the clockwise direction. (c) Distributions of Ising interactions $J_{ij}$ [Eq.~(\ref{eq_J})]. (d) Average activities $\langle x_i\rangle$ versus local fields $h_i$, where each point represents one neuron, and the dashed line illustrates the independent prediction [Eq.~(\ref{indmeans})]. 
\label{fig_model}}
\end{figure}

Applying our method, we identify the tree of maximally informative correlations [Fig.~\ref{fig_model}(a)], which captures $I_\mathcal{T} = 26.2\,\text{bits}$ of information; this is more than $50\times$ the average we find on a random tree [Fig.~\ref{fig_model}(b)]. Although a tree includes only $2/N \approx 0.1\%$ of all pairwise correlations, we have $I_\mathcal{T} \approx 0.144 S_{\rm ind}$, so that the correlation structure we capture is as strong as freezing the states of $214$ randomly selected neurons. Another way to assess the strength of the interactions is to see that on the optimal tree the mean activity of each neuron $\langle x_i\rangle$ deviates strongly from what is predicted by the field $h_i$ alone, 
\begin{equation}
\langle x_i \rangle_{J = 0} = {1\over{1 + e^{-h_i}}}.
\label{indmeans}
\end{equation}
This is shown in Fig.~\ref{fig_model}(d), where we compare the optimal tree with a random tree. So despite limited connectivity, the effective fields
\begin{equation}
h_i^{\rm eff} = h_i + \sum_j J_{ij}\sigma_j
\end{equation}
are very different from the intrinsic biases $h_i$. 


In the model of Eq.~(\ref{eq_PG}), positive (negative) $J_{ij}$ means that activity in neuron $i$ leads to activity (silence) in neuron $j$. For random trees, the interactions $J_{ij}$ are split almost evenly between positive and negative [Fig.~\ref{fig_model}(c)]; this distribution of interactions is consistent with previous investigations of systems of $N\sim 100$ neurons where we can estimate and match all of the pairwise correlations \cite{Schneidman-01, Tkacik-03, Meshulam-02}. But since the largest mutual information are associated with positive correlations [Fig.~\ref{fig_data}(c)], the maximally informative tree produces strong interactions that are almost exclusively positive and quite large [Fig.~\ref{fig_model}(c)]. We have arrived, perhaps surprisingly, at an Ising ferromagnet. 

\begin{figure}[t]
\centering
\includegraphics[width = .9\columnwidth]{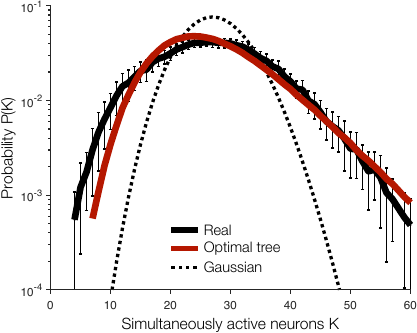} 
\caption{Predicting synchronized activity. Distribution $P(K)$ of the number of simultaneously active neurons $K$ in the data (black), predicted by the maximally informative tree (red), and the Gaussian distribution for independent neurons with mean and variance $\langle K\rangle_\text{exp}$ (dahsed). To estimate probabilities $P(K)$ and error bars (two standard deviations), we first split the experiment into $1$--minute blocks to preserve dependencies between consecutive samples. We then randomly select one third of these blocks and repeat $100$ times. For each subsample of the data, we compute the optimal tree $\mathcal{T}$ and predict $P(K)$ using a Monte Carlo simulation of the model $P_\mathcal{T}$. \label{fig_synchrony}}
\end{figure}

Is it possible that a backbone of ferromagnetic interactions captures some of the collective behavior in the network? One signature of this collective behavior is the probability $P(K)$ that $K$ out of the $N$ neurons are simultaneously active within a window of time \cite{Schneidman-01, Tkacik-03, Lynn-04, Meshulam-02}. For independent neurons, this distribution approximately Gaussian at large $N$ (Fig.~\ref{fig_synchrony}, dashed), but even in relatively small populations we see strong deviations from this prediction, with both extreme synchrony (large $K$) and near silence (small $K$) much more likely than expected from independent neurons \cite{Schneidman-01}; this effect persists in the $N\sim 1500$ neurons studied here (Fig.~\ref{fig_synchrony}, black). The optimal tree captures most of this structure, correctly predicting $\sim 100\times$ enhancements of the probability that $K\sim 50$ or more neurons will be active in synchrony (Fig.~\ref{fig_synchrony}, red). Although the detailed patterns of activity in the system are shaped by competing interactions that are missing from our ferromagnetic backbone, this shows that large--scale synchrony can emerge from a sparse network of the strongest positive correlations.

\begin{figure}[b]
\centering
\includegraphics[width = \columnwidth]{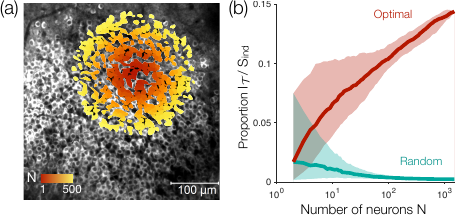} 
\caption{Scaling with population size. (a) Illustration of our growth process superimposed on a fluorescence image of the $N = 1485$ neurons in the mouse hippocampus. Starting from a single neuron $i$, we grow the population of $N$ neurons closest to $i$ (red to yellow). We then repeat this process starting from each of the different neurons. (b) Fraction of the independent entropy $I_\mathcal{T}/S_\text{ind}$ explained by the optimal and random trees as a function of population size $N$. Lines and shaded regions represent median and interquartile range over different central neurons. \label{fig_scaling}}
\end{figure}

Thus far, we have focused on a single population of $N \sim 1500$ neurons. But as we observe larger populations, how do the maximally informative correlations scale with $N$? To address this question, we build populations of increasing size by starting from a single cell and drawing concentric circles of increasing radii (in the spirit of Ref.~\cite{Meshulam-02}), then repeating this process starting from each of the different neurons [Fig.~\ref{fig_scaling}(a)]. This construction exploits the fact that the neurons in this region of the hippocampus lie largely in single plane. As the population expands, the independent entropy $S_{\rm ind}$ necessarily increases linearly with $N$ on average. The entropy of any tree model $S_\mathcal{T}$, which is an upper bound on the true entropy, is reduced by the total information $I_\mathcal{T} = S_\text{ind} - S_\mathcal{T}$. In Fig.~\ref{fig_scaling}(b) we see that the fractional reduction $I_\mathcal{T}/S_{\rm ind}$ grows slowly with $N$ for the optimal trees, while on random trees this fraction decays rapidly toward zero. 

We can understand the decay of fractional information in random trees because the mutual information between two neurons declines, on average, with their spatial separation, although there are large fluctuations around this average. These fluctuations mean that a tree built by connecting nearest spatial neighbors will be better than random but still substantially suboptimal, as will be explored elsewhere. In contrast, as we consider larger populations we uncover more and more of the large mutual information seen in the tail of Fig.~\ref{fig_data}(a), and this allows $I_\mathcal{T}/S_\text{ind}$ to increase with $N$ on the optimal tree [Fig.~\ref{fig_scaling}(b)]. There is no sign that this increase is saturating at $N\sim 10^3$, suggesting that our minimax entropy framework may become even more effective for larger populations.

In summary, it has been appreciated for nearly two decades that the maximum entropy principle provides a link from data directly to statistical physics models, and this is useful in networks of neurons as well as other complex systems \cite{Schneidman-01, Nguyen-01, Meshulam-17, Meshulam-02, Lezon-01, weigt+al_09, marks2011protein, lapedes2012using, Bialek-01, russ2020evolution, Lynn-04}. Less widely emphasized is that we do not have ``the'' maximum entropy model, but rather a collection of possible models depending on what features of the system behavior we choose to constrain. Quite generally we should choose the features that are most informative, leading to the minimax entropy principle \cite{Zhu-01}. As we study larger and larger populations of neurons we enter an undersampled regime in which selecting a limited number of maximally informative features is not only conceptually appealing but also a practical necessity.

The problem is that the minimax entropy principle is intractable in general. Here we have made progress in two steps. First, following previous successes, we focus on constraining the mean activity and pairwise correlations. Second, we take the lesson of the Bethe lattice and select only pairs that define a tree. Once we do this, the relevant statistical mechanics problem is solved exactly, and the optimal tree can be found in quadratic time \cite{Chow-01, Nguyen-01}. This means that there is a non--trivial family of statistical physics models for large neural populations that we can construct {\em very} efficiently, and it is worth asking whether these models can capture any of the essential collective behavior in real networks. We find that the optimal tree correctly predicts the distribution of synchronous activity (Fig.~\ref{fig_synchrony}), and these models capture more of the correlation structure as we look to larger networks [Fig.~\ref{fig_scaling}(b)]. The key to this success is the heavy--tailed distribution of mutual information [Fig.~\ref{fig_data}(a)], and this in turn may be grounded in the heavy--tailed distribution of physical connections \cite{Lynn-13}. While these models cannot capture all aspects of collective behavior, these observations provide at least a starting point for simplified models of the much larger systems now becoming accessible to experiments. \\

\begin{acknowledgments}
We thank L.\ Meshulam and J.L.\ Gauthier for guiding us through the data of Ref.~\cite{Gauthier-01}, and C.M.\ Holmes and D.J.\ Schwab for helpful discussions. This work was supported in part by the National Science Foundation, through the Center for the Physics of Biological Function (PHY--1734030) and a Graduate Research Fellowship (C.M.H.); by the National Institutes of Health through the BRAIN initiative (R01EB026943); by the James S McDonnell Foundation through a Postdoctoral Fellowship Award (C.W.L.); and by  Fellowships from the Simons Foundation and the John Simon Guggenheim Memorial Foundation (W.B.).
\end{acknowledgments}

\bibliography{MaxEntBib}

\end{document}